\documentstyle[prl,aps,epsfig,wrapfig,preprint,here]{revtex}
\input{epsf}                                                               
\tightenlines
\begin{document}                                           

\draft

\title{\bf {Direct measurement of the pion valence quark momentum
distribution, the pion light-cone wave function squared}}
%{\large\bf {Measurement of the pion light-cone  wave function}}
%%\author{E791 Collaboration}
% E791 author list Sep 98
%
% NOTE: This is a ReVTeX form.
%%%%%%%%%%%%%%%%%%%%   E791 Name list   %%%%%%%%%%%%%%%%%%%%%%%%%%%%%%%
%
\author{
    E.~M.~Aitala,$^9$
       S.~Amato,$^1$
    J.~C.~Anjos,$^1$
    J.~A.~Appel,$^5$
       D.~Ashery,$^{14}$
       S.~Banerjee,$^5$
       I.~Bediaga,$^1$
       G.~Blaylock,$^8$
    S.~B.~Bracker,$^{15}$
    P.~R.~Burchat,$^{13}$
    R.~A.~Burnstein,$^6$
       T.~Carter,$^5$
 H.~S.~Carvalho,$^{1}$
  N.~K.~Copty,$^{12}$
    L.~M.~Cremaldi,$^9$
 C.~Darling,$^{18}$
       K.~Denisenko,$^5$
       S.~Deval,$^3$
       A.~Fernandez,$^{11}$
       G.~F.~Fox,$^{12}$
       P.~Gagnon,$^2$
       S.~Gerzon,$^{14}$
       C.~Gobel,$^1$
       K.~Gounder,$^9$
     A.~M.~Halling,$^5$
       G.~Herrera,$^4$
 G.~Hurvits,$^{14}$
       C.~James,$^5$
    P.~A.~Kasper,$^6$
       S.~Kwan,$^5$
    D.~C.~Langs,$^{12}$
       J.~Leslie,$^2$
       J.~Lichtenstadt,$^{14}$
       B.~Lundberg,$^5$
       S.~MayTal-Beck,$^{14}$
       B.~Meadows,$^3$
 J.~R.~T.~de~Mello~Neto,$^1$
    D.~Mihalcea,$^{7}$
    R.~H.~Milburn,$^{16}$
 J.~M.~de~Miranda,$^1$
       A.~Napier,$^{16}$
       A.~Nguyen,$^7$
  A.~B.~d'Oliveira,$^{3,11}$
       K.~O'Shaughnessy,$^2$
    K.~C.~Peng,$^6$
    L.~P.~Perera,$^3$
    M.~V.~Purohit,$^{12}$
       B.~Quinn,$^9$
       S.~Radeztsky,$^{17}$
       A.~Rafatian,$^9$
    N.~W.~Reay,$^7$
    J.~J.~Reidy,$^9$
    A.~C.~dos Reis,$^1$
    H.~A.~Rubin,$^6$
 D.~A.~Sanders,$^9$
 A.~K.~S.~Santha,$^3$
 A.~F.~S.~Santoro,$^1$
       A.~J.~Schwartz,$^{3}$
       M.~Sheaff,$^{17}$
    R.~A.~Sidwell,$^7$
    A.~J.~Slaughter,$^{18}$
    M.~D.~Sokoloff,$^3$
       J.~Solano,$^{1}$
       N.~R.~Stanton,$^7$
      R.~J.~Stefanski,$^5$ 
      K.~Stenson,$^{17}$
    D.~J.~Summers,$^9$
 S.~Takach,$^{18}$
       K.~Thorne,$^5$
    A.~K.~Tripathi,$^{7}$
       S.~Watanabe,$^{17}$
 R.~Weiss-Babai,$^{14}$
       J.~Wiener,$^{10}$
       N.~Witchey,$^7$
       E.~Wolin,$^{18}$
     S.~M.~Yang,$^{7}$
       D.~Yi,$^9$
       S.~Yoshida,$^{7}$
       R.~Zaliznyak,$^{13}$
       and
       C.~Zhang$^7$ \\
(Fermilab E791 Collaboration)
}

\address{
$^1$ Centro Brasileiro de Pesquisas F\'\i sicas, Rio de Janeiro, Brazil\\
$^2$ University of California, Santa Cruz, California 95064\\
$^3$ University of Cincinnati, Cincinnati, Ohio 45221\\
$^4$ CINVESTAV, Mexico\\
$^5$ Fermilab, Batavia, Illinois 60510\\
$^6$ Illinois Institute of Technology, Chicago, Illinois 60616\\
$^7$ Kansas State University, Manhattan, Kansas 66506\\
$^8$ University of Massachusetts, Amherst, Massachusetts 01003\\
$^9$ University of Mississippi-Oxford, University, Mississippi 38677\\
%%$^{10}$ The Ohio State University, Columbus, Ohio 43210\\
$^{10}$ Princeton University, Princeton, New Jersey 08544\\
$^{11}$ Universidad Autonoma de Puebla, Mexico\\
$^{12}$ University of South Carolina, Columbia, South Carolina 29208\\
$^{13}$ Stanford University, Stanford, California 94305\\
$^{14}$ Tel Aviv University, Tel Aviv, Israel\\
%%$^{16}$ 317 Belsize Drive, Toronto, Canada\\
$^{15}$ Box 1290, Enderby, BC, VOE 1V0, Canada\\
$^{16}$ Tufts University, Medford, Massachusetts 02155\\
$^{17}$ University of Wisconsin, Madison, Wisconsin 53706\\
$^{18}$ Yale University, New Haven, Connecticut 06511\\
}

%\vspace{0.5cm}                                                                    

%~\\                                                                       
\date{\today}                                                        
                                                                           
\maketitle

\begin{abstract}
We present the first direct measurements of the pion valence quark
momentum distribution which is related to the square of the pion
light-cone wave function. The measurements were carried out using data on
diffractive dissociation of 500 GeV/c $\pi^-$ into di-jets from a platinum
target at Fermilab experiment E791. The results show that the $|q\bar
{q}\rangle $ light-cone asymptotic wave function, which was developed
using perturbative QCD methods, describes the data well for $Q^2
\sim 10 ~{\rm (GeV/c)^2}$ or more. 
We also measured the transverse momentum distribution of the diffractive
di-jets.
\end{abstract}
 
\newpage
%\vglue 0.5cm
%\noindent
%\section{Introduction}

The internal momentum distributions of valence quarks in hadrons
enter the calculation of a large variety of processes such as electroweak
decays, diffractive processes, meson production in $e^+e^-$ and $\gamma
\gamma$ annihilation, relativistic heavy ion collisions, and many others
\cite{stan1}. The momentum distribution amplitudes are generated from
the valence light-cone wave functions integrated over $k_t < Q^2$, where $k_t$
is the intrinsic transverse momentum of the valence constituents and
$Q^2$ is the total momentum transfer squared (Eqn. \ref{mdj}). Because of the
close relationship between
the two, the distribution amplitudes are often referred to as the
light-cone wave functions \cite{stesto}. Even though these amplitudes 
were calculated about 20 years ago, there have been no direct measurements
until those reported here. Observables which are related to these
distributions, such as the pion electromagnetic form factors, are rather
insensitive to the light-cone wave functions. \\

The pion wave function can be expanded in terms of Fock states:

\begin{equation}
\Psi = a_1 |q\bar {q}\rangle + a_2 |q\bar {q}g\rangle +
           a_3|q\bar {q}gg\rangle + \cdots\, .
\end{equation}
The first (valence) component is dominant at large 
$Q^2$. The other terms are suppressed by powers of $1/Q^2$ for each
additional parton, according to counting rules \cite{stesto,bf}. In
contrast, parton
distribution functions are inclusive momentum distributions of partons in
all Fock states. Here we are concerned with the momentum distribution of
only the valence quark-antiquark part. \\

Two functions have been proposed to describe the momentum distribution
amplitude for the quark and antiquark in the $|q\bar {q}\rangle$
configuration. The asymptotic function was calculated  using perturbative QCD
(pQCD) methods \cite{lb,er,bbgg}, and is the solution to the pQCD
evolution equation for very large $Q^2$ ($Q^2 \rightarrow \infty$):

\begin{equation}
\phi_{as}(x) =\sqrt{3} x(1-x).
\label{asy}
\end{equation}

\noindent
$x$ is the fraction of the longitudinal momentum of the pion
carried by the quark in the infinite momentum frame. The antiquark carries
a  fraction ($1 ~- ~x$). Using QCD sum rules, Chernyak and Zhitnitsky (CZ)
proposed \cite{cz} a function that is expected to be correct for low
Q$^2$:

\begin{equation}
\phi_{cz}(x) =5\sqrt{3} x(1-x)(1-2x)^2.
\label{cz}
\end{equation}
%\end{enumerate}

\begin{figure}[h]
%\begin{wrapfigure}{r}{8cm}
%\centerline{\epsfxsize=10cm \epsfbox{../figures/x_mc_wfsq_bw.eps}}
\centerline{\epsfxsize=10cm \epsfbox{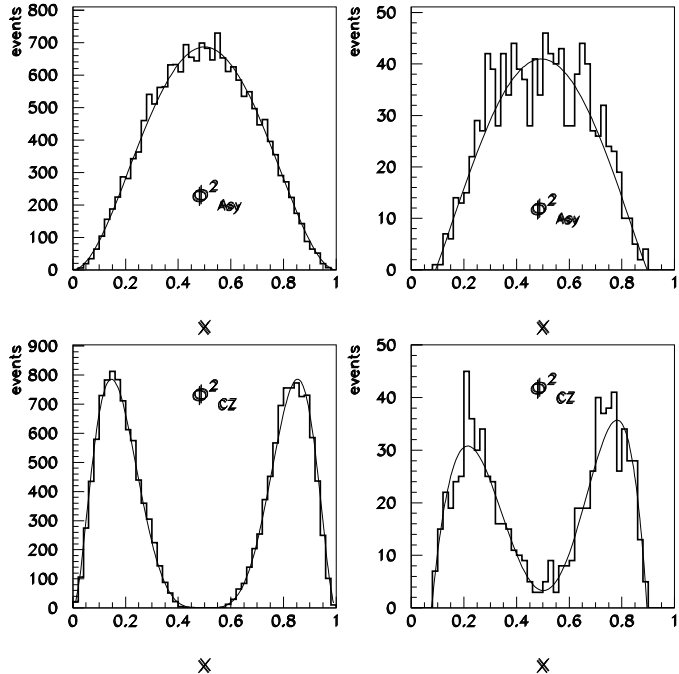}}
\caption{Monte Carlo simulations of squares of the two wave functions at the
quark  level (left) and of the reconstructed distributions of di-jets as
detected  (right).  $\phi_{Asy}^2$ is the asymptotic function (squared) and
$\phi_{CZ}^2$ is the Chernyak-Zhitnitsky  function (squared).
The di-jet mass used in the simulation is 6 GeV/c$^2$
and the plots are for $1.5 ~\rm{GeV/c} ~\leq ~k_t ~\leq ~2.5
~\rm{GeV/c}$.}
\label{fig:x_mc}
%\end{wrapfigure}
\end{figure}

%The predicted distribution amplitudes are plotted in Fig. \ref{fig:pionwf}
As can be seen from Eqns. \ref{asy} and \ref{cz}, and from Fig.
\ref{fig:x_mc}, there is a large difference between the two functions.
Measurements of the electromagnetic form factors of the pion were
considered to be the best
way to study these wave functions. A comprehensive summary of the status
of these measurements was published recently \cite{stesto}. Both existing
methods of measuring the pion electric form factor suffer from major
drawbacks. Those done using pions elastically scattered from atomic
electrons measure the  pion electric form factor only for very low
$Q^2$. The alternative method is to measure the electron-pion quasi-free
%$p(e,e' \pi^+)n$ 
scattering cross section \cite{stesto}. The
relation between the form factor and the longitudinal cross section is
model-dependent as it includes the $p \rightarrow n\pi^+$ matrix element.
Finally, the form factor is related to the integral over
the wave function and the scattering matrix element, reducing the
sensitivity to the wave function. Indeed, as shown in \cite{stesto} both
wave functions can be made to agree with the experimental data. A similar
situation exists for inelastic form factors, decay modes of heavy mesons,
etc.  Recent model-dependent analyses of CLEO data on meson-photon
transition form factors \cite{cleo,rad} are consistent with the asymptotic
wave function. The problem is that comparisons to these observables are
not sensitive to details of the wave function and thus cannot provide
critical tests of their $x$-dependence. Another open question is what
can be considered to be high enough $Q^2$ to qualify for perturbative QCD
calculations, what is low enough to qualify for a treatment
based on QCD sum rules and how to handle the evolution from low to 
high $Q^2$.\\

%\begin{wrapfigure}{r}{7cm}
%\begin{figure}[H]
%\centerline{\epsfxsize=6cm \epsfbox{pion_wf_p.eps}}
%\centerline{\epsfig{figure=../figures/wf_plt.ps,angle=90,width=10cm,height=7cm}}
%\epsfig{figure=pion_wf_p.eps,width=5cm,height=5cm}
%\vglue 0.5cm
%\caption{Two predictions of the pion wave function. The asymptotic
%function ($\phi_{Asy}$, dashed line) and the CZ function ($\phi_{CZ}$,
%solid line).}
%\label{fig:pionwf}
%\end{wrapfigure}
%\end{figure}

In this work we describe 
an experimental study that  maps the momentum distribution of the  
$ q $ and $ \bar q $ in the
$|q\bar {q}\rangle$ Fock state of the pion. This provides the first direct
measurement of the pion light-cone wave function (squared).
The concept of the measurement is the following: a high energy pion
dissociates diffractively on a heavy nuclear target imparting no energy 
to the target so that it does not break up. This is a coherent
process in which the quark and antiquark in the pion break apart and hadronize
into two jets. If in this fragmentation process each quark's momentum
is transferred to a jet, measurement of the jet momenta gives the
quark and antiquark momenta. Thus:

\begin{equation}
x_{measured} = \frac {p_{jet1}} {p_{jet1}+p_{jet2}}.
\label{x_meas}
\end{equation}

The diffractive dissociation of high momentum pions into two jets can be
described, like the inclusive Deep Inelastic Scattering
(DIS) and exclusive vector meson production
in DIS, by factoring out the perturbative high momentum transfer process
from  the soft nonperturbative part \cite{fact}. 
This factorization allows the use of common parameters to describe the
three processes. The {\em virtuality} of the process, $Q^2$, is given by
the mass-squared of the virtual photon in inclusive deep inelastic 
scattering (DIS) and in exclusive light vector meson production in DIS. In
exclusive DIS production of heavy vector mesons this mass is proportional to
that of the produced meson (e.g., $J/\psi$).
For diffractive dissociation into two jets, the mass-squared of the di-jets
plays this role. From simple kinematics and assuming that the masses
of the jets are small compared with the mass of the di-jets, the virtuality
and mass-squared of the di-jets are given by:

\begin{equation}
Q^2 \sim  M_{J}^2 = \frac{k_t^2}{x(1 - x)},
\label{mdj}
\end{equation}
where $k_t$ is the transverse momentum of each jet and reflects the intrinsic
transverse momentum of the valence quark or antiquark. By studying the momentum
distribution for various $k_t$ bins, one can observe changes in the apparent
fractions of asymptotic and Chernyak-Zhitnitsky contributions to the pion
wave function.

Fermilab experiment E791 \cite{e791} recorded $2\times 10^{10}$ events
from interactions of a 500 GeV/c  $\pi^-$ beam with carbon (C) and
platinum (Pt) targets. The trigger included a loose requirement on
transverse energy deposited in the calorimeters. Precision vertex and
tracking information was provided by 23 silicon microstrip detectors (6
upstream and 17 downstream of the targets), ten proportional wire-chamber
planes (8 upstream and 2 downstream of the targets), and 35 drift-chamber
planes. Momentum was measured using two dipole magnets. Two multicell, 
threshold \v{C}erenkov counters were used for $\pi$, $K$, and $p$
identification (not needed for this analysis). Only about 10\% of the
E791 data was used for the analysis presented here.
From these data, we selected interactions which were exclusive two jet events.
This focused on the jets as materializations of the valence quark and antiquark
in the pion.\\

The data were analysed by selecting events in which 90\% of the beam
momentum was carried by charged particles. This reduced the effects
of the unobserved neutral particles and
%, together with the high resolution of the microstrip detectors 
allowed for precise measurement of transverse
momentum. The selected events were subjected to the JADE jet-finding
algorithm \cite{jade}. The algorithm uses a cut-off parameter ($m_{cut}$)
whose value was optimized for this analysis using Monte Carlo simulation
studies in order to optimize the identification of di-jets. The di-jet
invariant mass was calculated 
assuming that all the particles were pions. To insure clean selection of 
two-jet events, a minimum $k_t$ of 1.25 GeV/c was required. Furthermore,
the di-jet nature of these events was verified by examining their relative
azimuthal angle, which for pure di-jets should be $180^{\circ}$. Strong peaking
at $180^{\circ}$ was observed (FWHM $\sim ~5^{\circ}$), and only events within 
$20^{\circ}$ of back-to-back were accepted for this analysis.\\ 

\begin{figure}[H]
%\begin{wrapfigure}{r}{8cm}
%\centerline{\epsfxsize=16cm \epsfbox{../figures/qt2_had97_15bw_cont_1.eps}}
\centerline{\epsfxsize=16cm \epsfbox{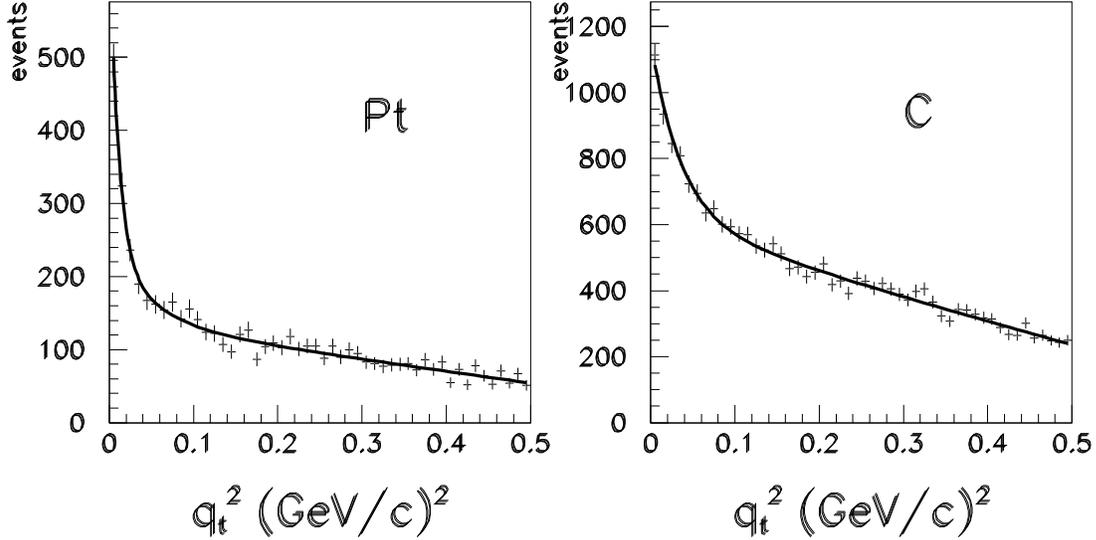}}
\vglue -7.0cm
\caption{$q_t^2$ distributions of di-jets with 1.25 GeV/c $\leq ~k_t$ from
interaction of 500 GeV/c $\pi^-$ with Carbon and Platinum targets.}
\label{data_diff}
%\end{wrapfigure}
\end{figure}

Diffractive di-jets were identified through the $e^{-bq_t^2}$ dependence
of their yield ($q_t^2$ is the square of the transverse momentum transferred
to the nucleus and $b = \frac{<R^2>}{3}$ where R is the nuclear radius).
Fig.~\ref{data_diff} shows the $q_t^2$ distributions
of di-jet events from platinum and carbon. The different 
slopes in the low $q_t^2$ coherent region reflect the different 
nuclear radii. Events in this region come from diffractive dissociation
of the pion. \\

The basic assumption that the momentum carried by the dissociating
$q \bar q$ is transferred to the di-jets was examined by Monte Carlo
(MC) simulation. MC samples with 4 and 6 GeV/c$^2$ mass di-jets
were generated with two different $x$ dependences at the quark level. 
The $x$-distributions were calculated by squaring the asymptotic and the
Chernyak-Zhitnitski (CZ) wave functions. One sample was simulated using 
the asymptotic wave function and the other, the CZ function.
The four samples were allowed to hadronize using the LUND PYTHIA-JETSET
\cite{mc} package and then passed through a simulation of the experimental 
apparatus to account for the effect of unmeasured neutrals and other
experimental distortions.

In Fig. \ref{fig:x_mc} the initial distributions at the
quark level are compared with the final distributions of the detected 
di-jets, including distortions in the hadronization process and 
influence due to experimental acceptance. As can be seen, the qualitative
features of the two distributions are retained. The results of this analysis
come from comparing the observed x-distribution to a combination of the
distributions shown, as examples, on the right of Fig. \ref{fig:x_mc}.\\

\begin{figure}[H]
%\centerline{\epsfxsize=14cm \epsfbox{../figures/fit_prl.ps}}
\centerline{\epsfxsize=14cm \epsfbox{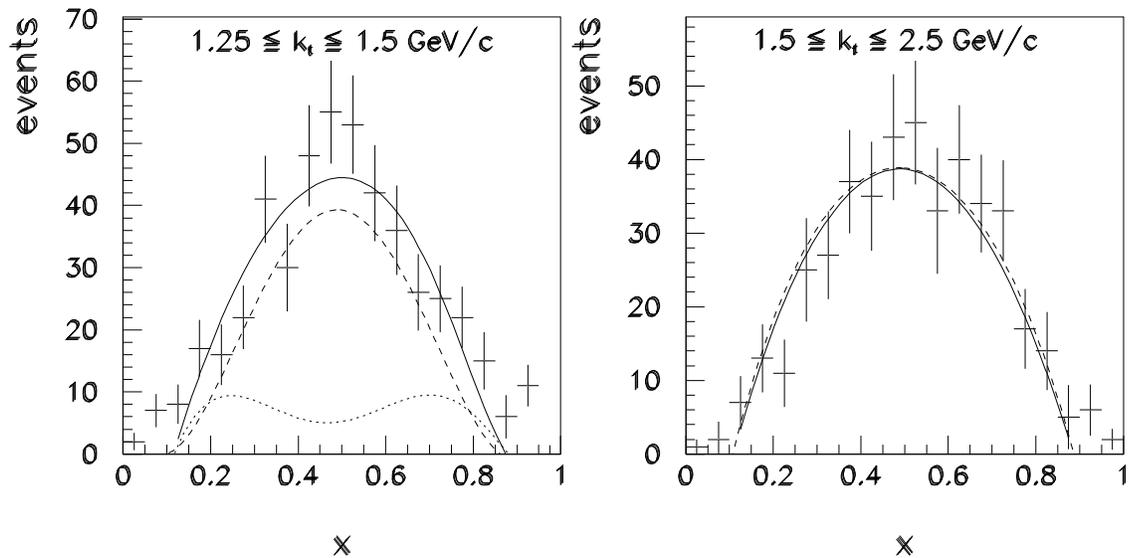}}
\vglue -9.0cm
\caption{The $x$ distribution of diffractive di-jets from
the platinum target for $1.25 \leq k_t \leq 1.5$ GeV/c (left) and for
$1.5 \leq k_t \leq 2.5$ GeV/c (right). The solid line is a fit to a
combination of the asymptotic and CZ wave functions. The dashed line shows
the contribution from the asymptotic function and the dotted line that of
the CZ function.} 
\label{xdatadif}
%\end{wrapfigure}
\end{figure}

For all results in this paper, we used data from the platinum target as 
it has a sharp diffractive distribution and a relatively low 
background. It is also expected that due to the color transparency effect
\cite{bbgg,fms,ct} this heavy target will better filter out  the high Fock
states. We used events with $q_t^2 ~< 0.015$ GeV/c$^2$. For these events, the 
value of $x$ was computed from the measured longitudinal momentum of each 
jet (Eqn. \ref{x_meas}). A background, estimated from the $x$
distribution for events with larger $q_t^2$, was subtracted.
This analysis was carried out in two windows of $k_t$:
$1.25 ~\rm{GeV/c} ~\leq ~k_t ~\leq ~1.5 ~\rm{GeV/c}$ and
$1.5 ~\rm{GeV/c} ~\leq  ~k_t ~\leq ~2.5 ~\rm{GeV/c}$. The experiment
data were compared to Monte Carlo simulations of di-jets having a mass of 4
GeV/c$^2$ for the lower window and 6 GeV/c$^2$ for the higher window. 
The resulting $x$ distributions are shown in Fig. \ref{xdatadif}.
In order to get a measure of the correspondence between the experimental
results and the calculated light-cone wave functions, we fit the results
with a linear combination of squares of the two wave functions. This
assumes an incoherent combination of the two wave functions and that the
evolution of the CZ function is slow (as stated in \cite{cz}). It is hard
to justify these two assumptions because it is hard to make model
independent evolutions
and to know the phase between the two amplitudes. We therefore regard this fit
as a qualitative indication of how well each function describes the data.
We use results of the simulated wave functions (squared) after they were
subjected to effects of experimental acceptance (Fig. \ref{fig:x_mc},
right). \\

\vglue -2.0cm
%\begin{wrapfigure}{r}{6cm}
\begin{table}
  \begin{center}
    \leavevmode
%    \begin{tabular}{|c|c|c|c|c|c|c|c|c|} %\hline
    \begin{tabular}{|c c c c c c c c c|} %\hline
$k_t$ bin &$a_{as}$ &$\Delta_{a_{as}}^{stat}$ & $\Delta_{a_{as}}^{sys}$ &
 $\Delta_{a_{as}}$ &  $a_{cz}$ & $\Delta_{a_{cz}}^{stat}$ &
$\Delta_{a_{cz}}^{sys}$ & $\Delta_{a_{cz}}$ \\
GeV/c & & & & & & & & \\\hline
1.25 - 1.5 &0.64 &$\pm0.12$&+0.07 -0.01& +0.14 -0.12 & 0.36 &$\mp0.12$
&-0.07 +0.01& -0.14 +0.12  \\ 
1.5 - 2.5 & 1.00  &$\pm0.10$& +0.00 -0.10 & +0.10 -0.14 &  0.00
&$\mp0.10$& -0.00 +0.10 &-0.10 +0.14
 \\ %\hline %\hline
    \end{tabular}
\vglue 0.5cm
\caption{Contributions from the asymptotic ($a_{as}$) and CZ ($a_{cz}$)
wave functions to a fit to the data.}
    \label{res}
  \end{center}
\end{table}
The measured $x$ distributions are shown in Fig. \ref{xdatadif} with the
combinations resulting from the above fits superimposed on the data. The
individual contributions from each wave function are shown as well. In addition
to the statistical errors of the fit, we considered systematic uncertainties
originating from the background subtraction, from the quality of the jets
and their identification and from using discrete-mass simulations. The
dominant contribution comes from the quality of the jets in the low $k_t$
region and from using discrete-mass MC in the high $k_t$ region. The 
results of the fits are given in Table \ref{res} 
in terms of the coefficients $a_{as}$ and $a_{cz}$ representing the
contributions of the asymptotic and CZ functions, respectively. The total
errors are obtained by adding the statistical and systematic errors in
quadrature.\\

The values of $\chi^2/dof$ were 1.5 and 1.0 for the low and high $k_t$
bins, respectively. The results for the higher $k_t$ window show clearly that
the asymptotic wave function describes the data very well. Because of the
dominance of the asymptotic wave function, this conclusion does not depend on
the assumptions made in fitting the data to a combination of the two
functions. The distribution in the lower
window is consistent with a significant contribution from the 
Chernyak-Zhitnitsky wave function. However, as stated above, if neither
function is dominant this can only indicate that at low  $k_t$ neither
function alone describes the data well. The requirement that  $k_t > $
1.5 GeV/c can be translated (Eqn. \ref{mdj}) to $Q^2 >\ \approx 10
%\sim 10 
~{\rm (GeV/c)^2}$. This shows that for these $Q^2$ 
values, the perturbative QCD approach that led to construction of the
asymptotic wave function is reasonable.\\

The $k_t$ dependence of diffractive di-jets depends on the quark
distribution amplitude. It was calculated recently by
Frankfurt, Miller, and Strikman \cite{fms}. They show that
the most important terms are those in which the $|q\bar{q}\rangle$ 
component of the pion interacts with two gluons emitted by
 the target. The predicted $k_t$ dependence 
%${d\sigma\over dk_t}$ $\sim {1\over k_t^6}$ to $ \sim {1\overk_t^7}$. 
can be seen from the cross section for this process \cite{fms}:

\begin{equation}
 {d\sigma\over dk_t^2} \propto
\left|\alpha_s(k_t^2) x_{Bj} G(x_{Bj},k_t^2)\right|^2
\left|\frac{\partial^2}{\partial k_t^2}\psi (x,k_t^2)\right|^2
\label{djcs}
\end{equation}

\noindent
where $\psi$ is the light-cone wave function.
Evaluation of the light-cone wave function at large $k_t$ as due to one gluon
exchange gives $\psi \sim \frac{\phi}{ k_t^2}$ with $\phi$ a slow function of
$k_t$ (e.g. the asymptotic  function). Given the weak
$\alpha_s(k_t^2)$ dependence, 
differentiating and squaring gives $\sim k_t^{-8}$. Since
$\alpha_s(k_t^2) G(x_{Bj},k_t^2) ~\sim ~k_t^{1/2}$ \cite{fms99} the expected
dependence of the cross section on $k_t$ is:

\begin{equation}
  {d\sigma\over dk_t} ~\sim ~k_t^{-6}.
\end{equation}

\begin{figure}[t]
%\centerline{\epsfxsize=16cm \epsfbox{../figures/kt_prl.eps}}
\centerline{\epsfxsize=16cm \epsfbox{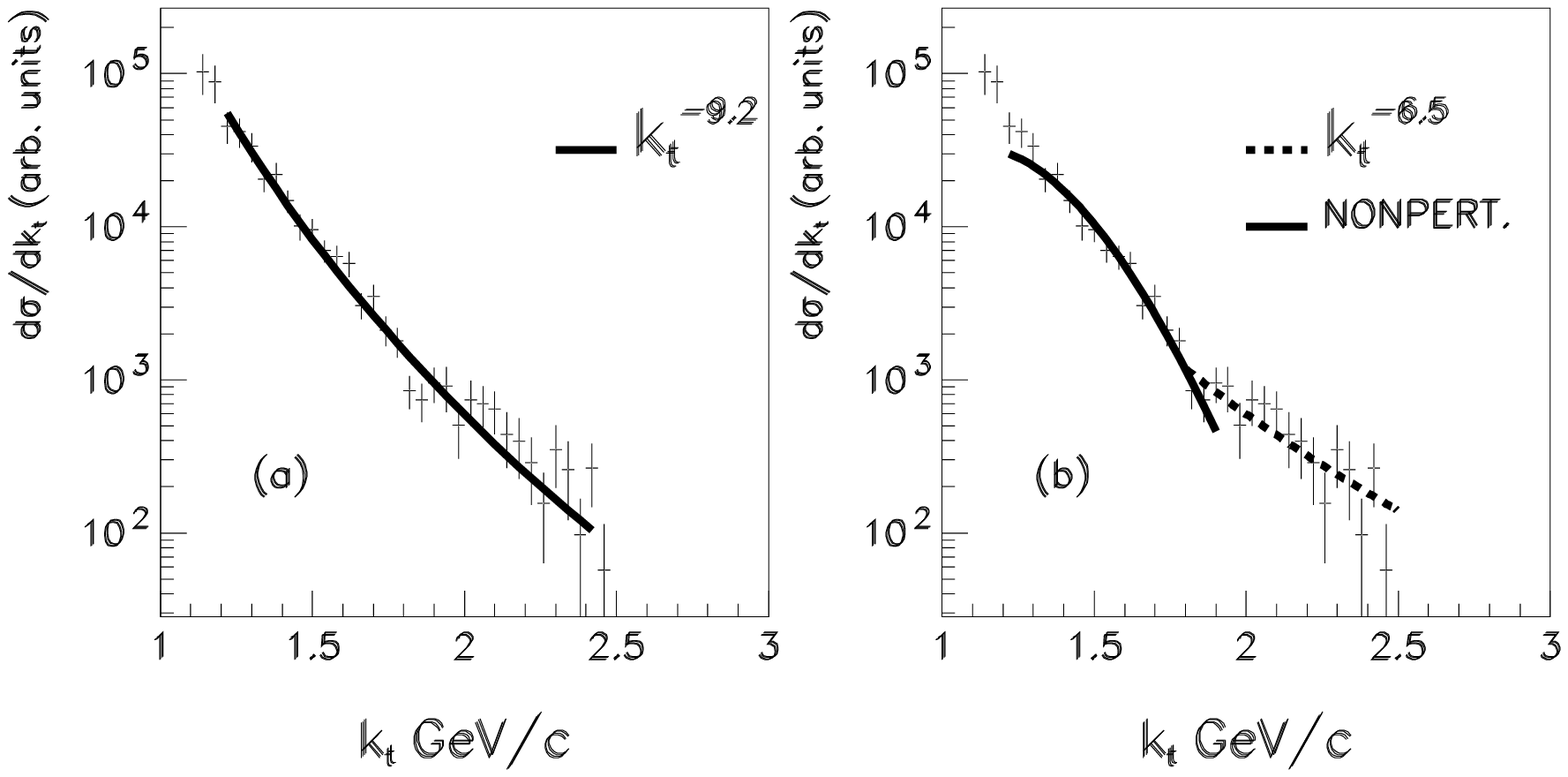}}
\vglue -6.0cm
\caption{Comparison of the $k_t$ distribution of acceptance-corrected data
with fits to
cross section dependence (a) according to a power law, (b) based on a 
nonperturbative Gaussian wave function for low $k_t$ and a power
law, as expected from perturbative calculations, for high $k_t$.}
\label{split}
\end{figure}

This prediction can be compared with the data. We use the MC simulations 
discussed above for di-jets having masses of 4, 5, and 6 GeV$/c^2$ and the
asymptotic wave function to correct for the experiment 
acceptance of the $k_t$ distribution. The corrected results are shown in
Fig. \ref{split}(a). Superimposed on the data is a power-law fit
$k_t^n$ for $k_t > 1.25$ GeV/c. We find n = $-9.2 \pm 0.4 (stat) \pm 0.3
(sys)$ with $\chi^2/dof$ = 1.0. This slope is significantly larger than
expected. We note, however,
that above $k_t \sim$ 1.8 GeV/c the slope changes (although the
statistical precision there is poor). A power law fit to this region (Fig.
\ref{split}(b)) results in $n = -6.5 \pm 2.0$ with $\chi^2/dof$ =
0.8, consistent with the predictions. This would support the evaluation of
the light-cone wave function at large $k_t$ as due to one gluon exchange.\\

The steep $k_t$-dependence in lower
$k_t$ region may be interpreted \cite{fms} as a manifestation of
non-perturbative
effects. We try the non-perturbative Gaussian function: $\psi \sim e^{-\beta
k_t^2}$ \cite{kro1}. When inserted in Eqn. \ref{djcs} and using
$\alpha_s(k_t^2) G(x_{Bj},k_t^2) ~\sim ~k_t^{1/2}$, we get:

\begin{equation}
 {d\sigma\over dk_t} = C(k_t^2 -2\beta k_t^4 + \beta^2 k_t^6)
e^{-2\beta k_t^2}
\end{equation}
where $C$ is a normalization factor.
In Fig. \ref{split}(b) we show a fit to this function in the low $k_t$
range yielding $\beta ~= ~1.78 ~\pm ~0.05 (stat) \pm 0.1 (sys)$ with
$\chi^2/dof$ = 1.1.
To the best of our knowledge, there is no previous report of a direct
measurement of $\beta$. Model-dependent values in
the range of 0.9 - 4.0 were used in calculations of the $\pi - \gamma$
transition form factors \cite{kro1}. The present results may indicate that
non-perturbative effects are noticeable up to $k_t ~\sim $ 1.5 GeV/c,
as is the case for the light-cone wave function which becomes
dominated by the asymptotic function only for larger $k_t$ values. \\

In summary, we have presented results of direct measurements of the
valence quark and antiquark 
momentum distributions in the pion. They show that above $k_t ~\sim$
1.5 GeV/c ($Q^2 ~\sim$ 10 GeV/c$^2$) the asymptotic distribution amplitude
calculated using perturbative QCD is applicable. The measured $k_t$ 
distribution in this region is also consistent with this conclusion.
In the lower $k_t$ region there may be other contributions, such as
from the CZ wave function or other nonperturbative effects.\\

We thank Drs. S.J. Brodsky, L. Frankfurt, G.A. Miller, and M. Strikman
for many fruitful discussions.
We gratefully acknowledge the staffs of Fermilab and of all the
participating institutions. This research was supported by the Brazilian
Conselho Nacional de Desenvolvimento Cient\'\i fico e Technol\'ogio,
the Mexican Consejo Nacional de Ciencia y Tecnologica,
the Israeli Academy of Sciences and Humanities, the
United States Department
of Energy, the U.S.-Israel Binational Science Foundation, and the United
States National Science Foundation. Fermilab is operated by the
Universities
Research Association, Inc., under contract with the United States
Department of Energy.


\begin{thebibliography}{99}

\bibitem{stan1} S.J. Brodsky, hep-ph/9908456, SLAC-PUB-8235.
\bibitem{stesto} G. Sterman and P. Stoler, Ann. Rev. Nuc. Part. Sci.      
{\bf 43}, 193 (1997), hep-ph/9708370.
\bibitem{bf}  S.J. Brodsky and G.R. Farrar,  Phys. Rev. Lett. {\bf 31}, 
1153 (1973).
\bibitem{lb} S.J. Brodsky and G.P. Lepage, Phys. Rev. D{\bf 22}, 2157
(1980);
         S.J. Brodsky and G.P. Lepage, Phys. Scripta {\bf 23}, 945 (1981);
         S.J. Brodsky, Springer Tracts Mod. Phys. 
         {\bf 100}, 81 (1982).
\bibitem{er} A.V. Efremov and A.V. Radyushkin, Theor. Math. Phys. {\bf 42}, 97
(1980).
\bibitem{bbgg} G. Bertsch, S.J. Brodsky, A.S. Goldhaber, and J. Gunion,
               Phys. Rev. Lett. {\bf 47}, 297 (1981). 
\bibitem{cz} V.L. Chernyak and A.R. Zhitnitsky, Phys. Rep. 
             {\bf112}, 173 (1984).
\bibitem{cleo}CLEO Collaboration, J. Gronberg {\it et al.} Phys. Rev. 
           D{\bf 57}, 33 (1998).
\bibitem{rad} P. Kroll and M. Raulfs, Phys. Lett. B{\bf 387}, 848 (1996);
I.V. Musatov and A.V. Radyushkin, Phys. Rev. D{\bf 56}, 2713 (1997).
\bibitem{fact} J.C. Collins, L.L. Frankfurt, and M. Strikman, Phys.~Rev.
D{\bf 56}, 2982 (1997).
\bibitem{e791}
%               J.~A.~Appel, Ann.~Rev.~Nucl.~Part.~Sci. 
%               {\bf 42},~367~(1992), and references therein;
% E.~M.~Aitala {\it et al.},~Phys.~Rev.~Lett.~{\bf 76},~364~(1996);
E791 Collaboration, E.M. Aitala {\it et al.}, EPJdirect {\bf C4}, 1
(1999).
\bibitem{jade}  
%CDF collaboration, FERMILAB-Conf-90/248-E; 
 JADE collaboration, W. Bartel et al., Z. Phys. {\bf C33}, 23 (1986).
\bibitem{mc} H.-U. Bengtsson and T. Sj\"{o}strand, Comp. Phys. Comm.
  ~{\bf 82}, 74 (1994); T. Sj\"{o}strand, PYTHIA 5.7 and JETSET 7.4
Physics and Manual, CERN-TH.7112/93, 1995, hep-ph/9508391.
%\bibitem{bbgg} G. Bertsch, S.J. Brodsky, A.S. Goldhaber, J. Gunion,
%                            Phys. Rev. Lett. {\bf 47}, 297 (1981)
\bibitem{fms}L.L. Frankfurt, G.A. Miller, and M. Strikman, Phys.
              Lett. {\bf B304}, 1 (1993).
\bibitem{ct}D. Ashery, hep-ex/0008036 and  E.M. Aitala {\it et al.},
         following letter, FERMILAB-PUB-00/220-E
\bibitem{fms99} L. Frankfurt, G.A. Miller, and M. Strikman, hep-ph/9907214,
Found. of Phys. {\bf 30}, 533 (2000).
%\bibitem{kro} T. Feldmann and P. Kroll Eur. Phys. Jour. {\bf C5}, 327
%(1998).
\bibitem{kro1} R. Jakob and P. Kroll, Phys. Lett. {\bf B315}, 463 (1993).

\end{thebibliography}
\end{document}